\documentstyle[12pt]{article}
\newcommand{\be}{\begin{equation}}
\newcommand{\ee}{\end{equation}}
\newcommand{\bea}{\begin{eqnarray}}
\newcommand{\eea}{\end{eqnarray}}

\input epsf
\begin{document}
\newcommand{\de}{\partial}

\title{Effective description of the LOFF phase of QCD}

\author{M.Mannarelli}

\maketitle
\begin{center}{\it Dipartimento di Fisica,
Universit\`a di Bari, I-70124 Bari, Italia  \\I.N.F.N.,
Sezione di Bari, I-70124 Bari, Italia\\
E-mail: massimo.mannarelli@ba.infn.it}
\end{center}

\begin{abstract}
We present an effective field theory for the crystalline
color superconductivity phase of QCD. It is kown that at high density and
at low temperature QCD 
exhibits a transition to a color superconducting phase characterized
by energy gaps in the fermion spectra. Under
specific circumstances the gap parameter  has a crystalline
pattern, breaking translational and rotational invariance. The
corresponding phase is the crystalline color superconductive
phase (or {\it LOFF} phase). We compute the parameters of the low
energy effective lagrangian describing the motion of the free
phonon in the high density medium and derive the phonon dispersion
law.
\end{abstract}

\section{Introduction}
In this talk We will discuss the recently proposed crystalline color superconducting phase
of QCD\cite{LOFF,nostro}.
This state is the QCD analogus to a state studied in QED by Larkin and Ovchinnikov\cite{LO}
and Fulde and Ferrell\cite{FF}. Therefore it has been named after them {\it LOFF} state.\\
It is known that at very high density and at low temperature quark matter exhibits color superconductivity\cite{review}.
Color superconductivity is a phenomenon analogus to the {\it BCS} superconductivity of QED.
According to the {\it BCS} theory, whenever there is 
an arbitrary small attractive interaction between electrons, the Fermi sea becomes unstable
with respect to the formation of bound pairs of electrons. Indeed the creation of a Cooper pair costs
no free energy, because each electron is on its own Fermi surface.
On the other hand there is an energy gain due to the attraction
of electrons. We call $\Delta$ the binding energy of each pair with respect to the Fermi level. It is possible to
show that the most energetically favored arrangement of electrons is made of two
electrons with opposite momenta and spins. That is the condensate has spin zero and momentum zero.
This is a normal superconductor. Now let us see what is a {\it LOFF} superconductor.

\section{{\it LOFF} phase in QED}
\begin{figure}[t]
\begin{center}
\vskip.6cm\epsfxsize=6truecm
\centerline{\epsffile{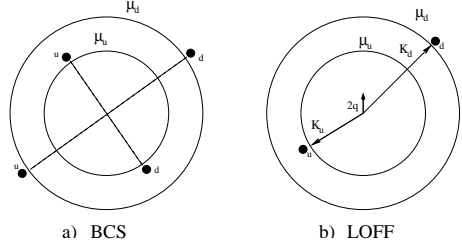}}
 \vskip1cm\noindent 
\caption{Fermi surfaces of down (d) and up (u) electrons (quarks). Black dots are electrons (quarks).
The {\it BCS} pair, in a), gains energy $\Delta$ but has $p^u_F=p^d_F$ and therefore pays free energy price.
Indeed one of the two electrons (quarks) of the pair has to stay far from its own Fermi surface.
The {\it LOFF} pair, in b), gains energy  $\Delta_{LOFF} \ll \Delta$ but does not pay any free energy price,
because each electron (quark) is on its own Fermi surface. For more details on figure b) see the text below.} 
\label{fig:1}
\end{center}
\end{figure}

Let us consider a system made of a ferromagnetic alloy containing paramagnetic impurities.
In first approximation the action of the impurities on the electrons may be viewed as a constant self-consistent exchange
field. Due to this field the Fermi surfaces of up and down electrons
(with respect to the direction of polarization of the medium) split, see figure (\ref{fig:1}).
The half-separation of the Fermi surfaces will be\cite{LO}
\be
I=N S a ~,
\label{I}\ee
where $N$ is the concentration of impurities, $S$ is their spin and $a$ is the integral of the exchange potential.
To determine the energetically favored state we have to compare the
{\it BCS} free energy with the free energy of the unpaired state. One has\cite{LO}:
\be 
F_{{\it BCS}}-F_{free} \propto I^2 - \Delta^2/2 ~. 
\label{energy}\ee
Therefore for $I > \Delta/\sqrt 2 $ the {\it BCS} state is no more energetically favorite.
For $I=\Delta/\sqrt 2 $ one expects that a first order phase transition
takes place from the {\it BCS} to the normal state. Or at least this is what one should naively
expect.
Instead it has been shown\cite{LO,FF} that for
\be 
\frac{\Delta}{\sqrt 2} < I < .75 \Delta ~,
\label{interval}\ee
it is energetically favorite a superconducting phase characterized by a condensate
of spin zero but with non zero total momentum.
We call $\Delta_{LOFF} \equiv \Delta_{LOFF}(I) $
the electron binding energy when the distance between the Fermi surfaces is $2 I$.
Increasing the half-distance between the Fermi spheres,
from $\frac{\Delta}{\sqrt 2}$ to $.75 \Delta$, the {\it LOFF} gap decreases to zero.
Therefore, for $I \simeq .75 \Delta $, a second order phase transition from
the {\it LOFF} phase to the normal phase takes place.  
In the {\it LOFF} phase pairs are made of
fermions which remain close to their own Fermi surfaces as shown in figure (\ref{fig:1}).
Therefore
the creation of
a pair costs no free energy.
The pair has total momentum $2 \vec q$, given by the formula
\be
\vec K_u + \vec K_d = 2 \vec q ~,
\label{2q}\ee 
where $\vec K_u$ and $\vec K_d$ are the momenta of up and down electrons.
The magnitude of $q$ is determined by minimizing the free energy of the {\it LOFF} state. 
The direction of $\vec q$ is chosen spontaneously by the condensate. We observe that not all electrons can 
condense in {\it LOFF} pairs. Only the electrons of a restricted region of the phase space can pair because of
equation (\ref{2q}).
Therefore $\Delta_{LOFF} \ll \Delta_{{\it BCS}}$
because of the reduced phase space available for coupling. Moreover $\Delta_{LOFF}$ turns out to be
a periodic function of the coordinates and we have a crystalline structure described by  $\Delta_{LOFF}(\vec r)$. 
  
\section{{\it LOFF} phase in QCD}
Let us consider quark matter at asymptotic density and at low temperature.
In the previous section we have learnt
that the {\it LOFF} phase  may arise when fermions
have different chemical potentials, and the potential difference
lies inside a certain window\cite{LOFF,LOFF2001,LOFF5,LOFFbis,LOFF7,LOFF10}.
Crystalline color
superconductivity is also expected to occur in case of different quark masses\cite{massdiff}.   
In what follows we shall consider matter made of massless quarks and electrons, in weak equilibrium and
electrically neutral. If we impose weak equilibrium we have 
$ \mu_u = \bar\mu -\frac{2}{3} \mu_e $ and $\mu_{d,s} = \bar\mu + \frac{1}{3} \mu_e $
where $\mu_{u,d,s,e}$ are quarks and electrons chemical potential. Requiring
electrical neutrality we get:
\be
\frac{2}{3} N_u -\frac{1}{3} N_d - \frac{1}{3} N_s - N_e = 0 ~,
\ee
where $N_{u,d,s,e}$ are the concentrations of quarks and electrons.
When $m_s=0$ we get $N_u=N_d=N_s$, $N_e=0$ and we expect matter to be in the color-flavor locked phase
(CFL) \cite{review}. For $m_s \sim \bar \mu$  we have  
$\mu_u = \bar{\mu} - \delta \mu ~, \mu_d = \bar{\mu} + \delta \mu$, and there are just two active flavors.

\section{Velocity dependent effective lagrangian}
In this Section we give a brief review of
the effective lagrangian approach for the {\it LOFF} phase\cite{LOFF5}, based on velocity-dependent fields.
We perform a Fourier transformation of the fermion field
 \be \psi(x)=\int\frac{d^4 p}{(2\pi)^4}e^{-ip\cdot x}\psi(p) ~,\ee
 and we decompose the fermion momentum as
 \be p^\mu_i=\mu_i v^\mu_i+\ell^\mu_i ~, \label{eq:2}\ee where
$i=1,2$ is a flavor index, $v^\mu_i=(0,\vec v_i)$ and $\vec v_i$ the Fermi velocity (for massless
fermions $|\vec v|=1$); finally $\ell^\mu_i$ is a residual momentum.
By the decomposition (\ref{eq:2}) only the positive energy
component $\psi_+$ of the fermion field survives in the lagrangian
in the $\mu\to\infty$ limit  while the negative energy
component $\psi_-$ can be integrated out. These effective fields
are velocity dependent and are related to the original fields by
\be
\psi(x)=\sum_{\vec v}e^{-i\mu v\cdot x
}\left[\psi_+(x)+\psi_-(x)\right]\label{decomposition}
 ,\ee where
\be\psi_\pm(x)=\frac{1\pm \vec \alpha\cdot \vec v }2 \int
\frac{d\vec\ell}{(2\pi)^3} \int_{-\infty}^{+\infty}
\frac{d\ell_0}{2\pi} \, e^{-i\ell\cdot x} \psi_{\vec v}(\ell)\
.\ee Here $\displaystyle\sum_{\vec v} $ means an average over the
Fermi velocities  and \be\psi_\pm(x)\equiv\psi_{\pm,\vec v}(x) \ee
are velocity-dependent fields.\\
Now we write the condensates. We have a scalar condensate (sc)
\be
< \epsilon_{ij}\epsilon_{\alpha\beta 3 } \psi^{i\alpha}( \vec
x)C\psi^{j\beta}(\vec x)> \propto \Delta_A e^{2i\vec q\cdot\vec
x}\label{scalar}\ \ee 
and a spin 1 vector condensate (vc)
\be <\sigma^1_{ij}\epsilon_{\alpha\beta 3 }
\psi^{i\alpha}(\vec x)C\sigma^{03}\psi^{j\beta}(\vec
x)> \propto \Delta_B e^{2i\vec q\cdot\vec x}\ .\label{vector}\ee 
The lagrangian for the (sc) is
\bea {\cal L
}^{(s)}_\Delta&=&-\frac{\Delta_{A}} 2 \,\sum_{\vec v_1,\vec v_2}
\exp\{i\vec x\cdot\vec\alpha(\vec v_1,\,\vec v_2,\,\vec q
)\}\epsilon_{ij}\epsilon^{\alpha\beta 3}\psi_{-\,\vec
v_i\alpha}(x)C \psi_{-\,\vec v_j\beta}(x)\cr && -(L\to
R)+{\rm h.c.}\ ,\label{loff6}\eea 
where $ \vec\alpha(\vec
v_1,\,\vec v_2,\,\vec q)=2\vec q-\mu_1\vec v_1-\mu_2\vec v_2\ $.
We take $\vec q = (0,0,q)$, therefore the components
of $\vec \alpha$ satisfy the equations
\bea 
\alpha_x&=&\ -\mu_1\sin\alpha_1 \cos\phi_1 -\mu_2\sin\alpha_2 \cos\phi_2 \ ,\cr
\alpha_y&=&\  -\mu_1\sin\alpha_1 \sin\phi_1 -\mu_2\sin\alpha_2 \sin\phi_2 \ ,\cr \alpha_z&=&\ 2\,q\,
-\,\mu_1\cos\alpha_1\,-\,\mu_2\cos\alpha_2 \ .\label{alfaz}\eea
In the limit $\mu_1,\mu_2 \rightarrow \infty$ we get
\bea 
\phi_2&\ =\ &\phi_1+\pi \\
\alpha_2&\ \equiv\ &\theta_q\ =\
\arccos\frac{\delta\mu}q\,+\, {\cal O}\left( \frac{\delta\mu}{\mu} \right),
\label{tetaq}\\
\alpha_1&\ =\ &\alpha_2+\pi + {\cal O}\left( \frac{\delta\mu}{\mu} \right) ~.  \eea
Therefore the velocities are almost opposite $ \vec v_1\simeq -\vec v_2\ $, and we have to deal with a no more
symmetric sum over velocities\cite{nostro}, because the angle $\alpha_2$ is fixed.

\section{Phonon quark interaction}
The condensates (\ref{scalar}) and (\ref{vector})
explicitly break rotations and translations. We have an
induced lattice structure given by parallel planes perpendicular
to $ \vec n = \vec q / |\vec q|$:\be \vec n \cdot\vec x\,=\,\frac{\pi k}{q}\hskip 1cm
(k\,=0,\, \pm 1,\,\pm 2,...) \ .\label{planes}\ee 
Lattice planes are allowed to fluctuate in two ways. We describe these fluctuations
by means of two fields $\phi$ and $R$:
\be \vec n \cdot\vec
x\,\to\, \vec n \cdot\vec x\,+\,\frac{\phi}{2qf} \equiv\, \frac{\Phi}{2q} ~,
\ee
\be
\vec n\,\to\,\vec R\ .\label{43}
\ee
But $\phi$ and $R$ are not independent, indeed it is possible to show\cite{nostro} that
\be
\vec R =\frac{\vec \nabla\Phi}{|\vec \nabla\Phi|}\, .\label{r3}
\ee 
Therefore there is just one independent degree of freedom, i.e. the phonon. 
\section{Effective lagrangian}
By a bosonization procedure\cite{nostro} it is possible to get the effective action for the phonon and
to calculate the polarization tensor. At the lowest order in the momentum $\vec p$
of the phonon we have
\be\Pi(0)=\Pi(0)_{s.e.}\,+\,\Pi(0)_{tad}\, =\,0\ ,
\label{masses}\ee
\be
\Pi(p)\,=\,-\,\frac{\mu^2}{4\pi^2 f^2}\Big[
 p_0^2-v_\perp^2(p_x^2+p^2_y)-v_\parallel^2 p^2_z
 \Big]\, ,\label{final2bis}\ee
where
\bea v_\perp^2&=&\frac 1
2\sin^2\theta_q+ {\cal O}\left(\frac{\Delta^{(v)}}q\right)^2 \, , \\
v_\parallel^2&=&\cos^2\theta_q
  \ , \eea
are the velocities perpendicular and parallel to $\vec q$.
From eq.(\ref{masses}) we see that the phonon is massless. 
From eq.(\ref{final2bis}) we have the dispersion law for the phonon:
\be E(\vec
p)=\sqrt{v_\perp^2(p_x^2+p^2_y)+v_\parallel^2 p^2_z}\ ~.\ee
In conclusion we can say that the dispersion law for the phonon is anisotropic in two ways.
Indeed the velocity of propagation of the
phonon in the plane perpendicular to $\vec q$, i.e. $ v_\perp$,
is not equal to $v_\parallel $, that is the velocity
of propagation of the phonon along $\vec q$. Moreover $p_z$ is a quasi-momentum.


\begin{thebibliography}{99}
%
\bibitem{LOFF}M. Alford, J. A. Bowers and K. Rajagopal,
Phys. Rev. {\bf D63} (2001) 74016.
%
\bibitem{nostro}
R.Casalbuoni, R. Gatto, M. Mannarelli and G. Nardulli, hep-ph/0201059.
%
\bibitem{LO}A. I. Larkin and Yu. N. Ovchinnikov, {\it Zh. Eksp.
Teor. Fiz.} {\bf 47}, 1136 (1964) ({\it Sov. Phys. JETP} {\bf 20}, 762
(1965)).
%
\bibitem{FF}P.Fulde and R. A. Ferrell, {\it Phys. Rev} {\bf 135}, A550 (1964).
%
\bibitem{review} For a review see K.Rajagopal and F.Wilczek,
in Handbook of QCD, M.Shifman ed. (World Scientific 2001), hep-ph/
0011333. For earliest papers on color superconductivity see B.
Barrois, Nucl. Phys. B {\bf 129}, 390 (1977); S. Frautschi,
Proceedings of workshop on hadronic matter at extreme density,
Erice 1978. See also: D. Bailin and A. Love, Phys. Rept. {\bf
107}, 325 (1984) and references therein.
%
\bibitem{LOFF2001}J. A. Bowers, J. Kundu, K. Rajagopal and E.
Shuster, Phys. Rev. D {\bf 64},  014024 (2001) hep-ph/0101067.
%
\bibitem{LOFF5}
R.Casalbuoni, R. Gatto, M. Mannarelli and G. Nardulli,
Phys. Lett {\bf B511},  (2001) 218.
%
\bibitem{LOFFbis}A. K. Leibovich, K. Rajagopal and E. Shuster,
 Phys. Rev. {\bf D64} (2001) 94005.
%
\bibitem{LOFF7}
D. K. Hong, Y. J. Sohn, hep-ph/0107003.
%
\bibitem{LOFF10}
D. K. Hong, hep-ph/0112028.
%
\bibitem{massdiff}
J. Kundu and K. Rajagopal, hep-ph/0112206.

\end{thebibliography}
\end{document}